\documentclass[journal]{IEEEtran}
\usepackage{amsmath}
\usepackage{url}
\usepackage{graphicx} 
\usepackage{balance}

\begin{document}

\title{``Estimating software project effort using analogies'':\\ Reflections after 28 years}

\author{Martin Shepperd% <-this % stops a space
\thanks{M. Shepperd is with the Department
of Computer Science, Brunel University of London, UB8 3PH, UK. Email: martin.shepperd@brunel.ac.uk.}% <-this % stops a space
}

\markboth{Transactions on Software Engineering,~Vol.~??, No.~??, January~2025}%
{Shepperd \MakeLowercase{\textit{et al.}}: Estimating Effort ...: Reflections after 28 years}

\maketitle

\begin{abstract}
Background: This invited paper is the result of an invitation to write a retrospective article on a ``TSE most influential paper'' as part of the journal's 50th anniversary.\\
Objective: To reflect on the progress of software engineering prediction research using the lens of a selected, highly cited research paper and 28 years of hindsight.\\
Methods: The paper examines (i) what was achieved, (ii) what has endured and (iii) what could have been done differently with the benefit of retrospection.\\
Conclusions: While many specifics of software project effort prediction have evolved, key methodological issues remain relevant. The original study emphasised empirical validation with benchmarks, out-of-sample testing and data/tool sharing. Four areas for improvement are identified: (i) stronger commitment to Open Science principles, (ii) focus on effect sizes and confidence intervals, (iii) reporting variability alongside typical results and (iv) more rigorous examination of threats to validity.  

\end{abstract}

\begin{IEEEkeywords}
Prediction; Project management; Effort prediction; Analogical reasoning; Empirical validation; Reproducibility.
\end{IEEEkeywords}

\section{Introduction}\label{Sec:intro}

\IEEEPARstart{T}his paper is the result of an invitation to contribute to the 50th anniversary of the journal.  I would like to thank the editors of IEEE Transactions on Software Engineering for this opportunity to write a reflective piece on the paper entitled ``Estimating software project effort using analogies'' \cite{Shepperd1997estimating} that I and Chris Schofield wrote back in 1997.

Back in the 1980s and 1990s a major challenge in software engineering was effective software project management, and within this, a particular difficulty was accurate prediction of resources, most notably costs and schedule.  Informal, subjective methods or `guestimates' often struggled resulting in costings and delivery dates that diverged wildly from reality \cite{Kraut1995coordination,Lederer1995causes}.  This in turn led to research and interest in more objective and data-driven prediction systems.  

Although the idea of estimating by analogy had been previously mooted by others, e.g., Boehm~\cite{Boehm1984software}, we proposed a formal implementation of the process of finding and using analogies or donor cases.  So in our research, we investigated the application of case-based reasoning (CBR) \cite{Kolodner1992introduction,Bergmann2005representation}, i.e., searching for and using analogies to support the prediction of software project effort. This was implemented by plotting completed projects, with known effort, in feature-space and then determining potential analogies for the target case, by minimising standardised Euclidean distance. Then the $k$ closest analogies were used as donors, with different strategies for pooling and adapting the various known completion effort values.  Typically, the number of donor analogies is in the range $2 \le k \le 5$ but is governed by factors such as the size of the case base $n$ and its homogeneity.

Analogy was seen as attractive because it is able to deal with poorly understood domains like software projects because solutions are based upon what has actually happened.  Additionally, since it is a form of lazy learning, there is only a requirement to solve \textit{actual} prediction problems as they arise and not for an entire domain of potential or hypothetical problems.

We then systematically evaluated the method over nine different data sets and used stepwise regression as a benchmark prediction method. In all cases, analogy was as good or better than the benchmark, which led us to conclude that it was a competitive software effort prediction method.

The remainder of this paper is organised as follows. Next, Section~\ref{Sec:achieved} considers, with the benefit of hindsight, what our original actually accomplished.  This is followed by a review in Section~\ref{Sec:persist} of what has, in some sense, been influential to the research and practitioner communities, and so has endured.  And then, perhaps unsurprisingly, a longer Section~\ref{Sec:different} of regrets and insights as to how the research `ought' to have been conducted deploying the benefits of 28 years of hindsight! Finally, in Section~\ref{Sec:lessons} the paper concludes with a list of more general lessons.

\section{Original research}\label{Sec:achieved}

The contributions of our 1997 study and paper fall into two categories, the direct findings themselves and  methodological issues.  

First then, the \textit{findings or results}.  As mentioned in the Introduction, software projects were growing rapidly in size, complexity and significance during the 1980s and 90s.  Yet, managing and controlling such projects was extremely challenging.  The idea of data or project metrics being collected was only slowly being adopted and developing formal, objective prediction systems still in its infancy.

Although various general-purpose cost models were being mooted, most notably Boehm's COCOMO \cite{Boehm1984software}, these tended to suffer from two serious drawbacks.  First, they required some kind of size input parameter, typically lines of code (LOC) which was seldom available until near the end of the development and second, independent studies e.g.,~\cite{Kemerer1987empirical} showed they performed very poorly outside their own development domain.  In other words, they did not generalise. 

In response, other researchers such as Kitchenham and Taylor~\cite{Kitchenham1985software} suggested collecting local data and using linear regression methods to fit a local model.  However, there was also interest in other methods that were perceived as being closer to how experts estimated, in particular the use of analogies, specifically similar completed projects for which costs or other values of interest were known.  Early work by Prietula et al.~\cite{Prietula1996software} suggested a hybrid approach using case-based reasoning technology, but with input and adaptation from experts.  In contrast, we pioneered purely algorithmic methods and implemented them in a CBR shell named ANGEL initially written in Visual Basic but subsequently extended, rewritten in Java and made freely available online.\footnote{Our ANGEL CBR tool was downloaded more than 10,000 times.}

Our paper also emphasised the need for explanatory value.
As we stated "[u]sers may be more willing to accept solutions from analogy based systems since they are derived from a form of reasoning more akin to human problem solving'' than from a black box. This is particularly so, since project prediction might be characterised as infrequent but high-value.   
    
Our generally positive results across nine diverse datasets, representing a broad range of software projects, inspired both other research groups and us to further develop this line of work.  More generally, the study contributed to the growing interest in deploying some of the nascent machine learning methods beyond CBR, e.g., rule induction methods and multilayer perceptrons.

In terms of \textit{methodological issues}, we contributed in a number of different ways.

First, we were unambiguous that when assessing prediction systems---as opposed to model fitting---it is essential to consider the expected performance on yet unseen data, the so-called out-of-sample (otherwise known as the hold-out set) loss or error.  This principle was already widely understood in the machine learning community, but perhaps less so by software engineering researchers.  Furthermore, given the increasingly complex nature of the algorithms, parameter tuning and data cleaning, this separation of training and test data could be unintentionally lost, e.g., by optimising feature subsets for all data \textit{prior} to splitting it into training and test (hold-out) subsets.

The method we deployed for hold-out was leave-one-out-cross-validation (LOOCV).\footnote{At the time we referred to LOOCV as a jackknife, which is indeed an identical procedure but the term is normally used for studying the bias or variance of a parameter rather than for assessing the generalisation error of prediction system.}  This process involves successively removing each case or software project from the training set, using it as the target for prediction, and then replacing it before moving on to the next case, continuing until all projects have been used as hold-out cases. The advantage of this approach to out-of-sample validation is that it is deterministic so the results will not depend upon chance vagaries, hence the results are easier to compare across studies.  On the other hand, as data set size increases LOOCV becomes more computationally demanding.

Second, and related to the first point, we applied a ``no peeking'' rule to maintain the integrity of early-stage predictions. Since our goal was to predict effort at the start of a project, we excluded information like LOC, which is typically unavailable until the project nears completion.

Third, we used a substantial number (9) of diverse data sets containing a total of 275 software projects.  The diversity was intended to promote the generalisability of our findings although this was done somewhat informally and without defining the intended population, a point to which we return in Section~\ref{Sec:different}.  All of these data sets were publicly available or alternatively published within the paper.  The idea of sharing data, and therefore being able to repeat and compare studies is now commonplace, though much less so back in the 1990s.  

Fourth, our study emphasised how predictive performance was to be assessed and the problems with some biased measures such as mean magnitude of relative error (MMRE) and Pred(25).\footnote{The prediction accuracy metrics MMRE and Pred(25) were popularised by Conte, Dunsmore and Shen~\cite{Conte1985software,conte1986software} in the 1980s. 
 MMRE is $100/n \sum^{i=n}_{i=1}(Y-\hat{Y})/Y$ where $n$ is the number of predictions and $Y$ the predicted commodity, in our case the project effort.  Pred(25) is the proportion of predictions that fall within the generally accepted, but somewhat arbitrary, threshold of 25\% of the true value. }  As we noted, the ``choice of accuracy measure to a large extent depends upon the objectives of those using the prediction system'' \cite{Shepperd1997estimating}.  Nevertheless, despite some concerns about the asymmetry of MMRE our analysis proceeded to use these two performance metrics on the grounds that they are ``widely used, thereby rendering our results more comparable to those of other workers''~\cite{Shepperd1997estimating}.  This is another point to which we return in Section~\ref{Sec:different}.  
    
Fifth, we performed a sensitivity analysis of the impact of the size of the training data set.  A practical and important question---particularly in a setting where project data could be scarce---was how much data was needed.  Since CBR is a lazy learning technology, we speculated that the need for data might be lessened.  To explore this, we simulated the growth of the training set over time.  Ideally, one would grow the data set in time order; however, project completion dates were not available, so we randomised the order (and repeated this procedure three times to expose the variability of the results).  From the analysis of two data sets, the results suggested accuracy grows with training set size and usefully that it started to stabilise from approximately 10 projects, suggesting that predictions made with less data are likely to be high risk.

\section{Legacy and Influence}\label{Sec:persist}

Our pioneering use of a formal implementation of analogical reasoning for prediction (otherwise case-based reasoning) has led to its widespread application plus its further development and refinement.  There are systematic reviews by ourselves \cite{Mair2005consistency} which locates 20 relevant studies and more recently Idri et al.~\cite{Idri2015analogy} who cover 65 studies published up to 2012.  These give some idea of the extent of research into analogy-based project prediction.  Interestingly, both reviews point to mixed evidence regarding whether analogy or a benchmark method such as regression analysis is `better'. 

\begin{table}[!ht]
    \centering
    \caption{Summary of systematic review results comparing Analogy and Regression-based prediction}
    \begin{tabular}{|l|r|r|r|l|}
    \hline
        \textbf{Review} & \textbf{Results+} & \textbf{Results-} & \textbf{Results=} & \textbf{Unit} \\ \hline
        Mair~\cite{Mair2005consistency}  & 9 & 7 & 4 & Paper \\ \hline
        Idri~\cite{Idri2015analogy} & 36 & 4 & 0 & Study x dataset \\ \hline
    \end{tabular}
    \label{TabSysRevs}
\end{table}

Table~\ref{TabSysRevs} summarises the empirical comparisons between analogy and regression-based effort prediction where Results+ indicates evidence in favour of analogy, Results- is evidence in favour of regression analysis and Results= indicates no strong evidence either way.  Note that the two systematic reviews utilise different units of analysis, which may explain the slightly inconsistent results.  Overall, we can conclude that there is some independent, supportive evidence suggesting that analogy-based prediction is a competitive method compared to the benchmark of linear regression analysis.  However, the evidence is not fully consistent, not least because of differences between data sets, configuration of the prediction methods and variations in experimental set up.

One problem that has been tackled subsequent to the 1997 paper is how to select appropriate subsets of case or analogy features.  It became apparent that simply using all features, especially if equally weighted, was generally suboptimal.  Indeed, this problem, known as feature subset selection, is endemic to machine learning prediction and to modelling in general \cite{John1994irrelevant}.  Different approaches include both statistical \cite{Li2009study} and search-based methods \cite{kirsopp2002search} with the general consensus that this is an extremely important but frequently computationally challenging aspect of analogy-based prediction.

Other researchers have built upon our work by exploring methods to improve similarity assessment (Keung et al.~\cite{Keung2008analogy}) and donor selection strategies (Kocaguneli et al.~\cite{Kocaguneli2011exploiting}).   

Another aspect that has been developed and extended is the area of adaptation.  The idea is that history repeats itself but not \textit{exactly}, hence some research has explored adapting the solutions proposed by donor cases to better fit the new target problem.  Such work includes \cite{Kirsopp2003empirical} deploying some simple linear methods. A later empirical analysis by Phannachitta et al.~\cite{Phannachitta2017stability} found that surprisingly, the simple linear methods tended to outperform the more sophisticated methods.

Three other directions that have been explored to some extent are based on the use of fuzzy features where the idea is that fuzzy features capture the inherent uncertainty of project measurement better than crisp features \cite{Idri2016improved} and Grey Relational Analysis (GRA) \cite{Song2005using,Song2011predicting}.  Both CBR and GRA evaluate the similarity between cases or projects, however, GRA handles problems like missingness and uncertainty more directly.  Some initial results were encouraging but GRA has yet to be widely deployed.  A third proposal is based on the idea of stacking and using multiple models \cite{Kaushik2022stacking}.

To summarise, using analogy-based prediction in software engineering has been widely deployed.  To quote Phannachitta et al., ``effort estimation based on Case-based reasoning is one of the most adopted methods in both the industry and research communities"~\cite{Phannachitta2017stability}.  It has spawned a good deal of creativity and new approaches to prediction and continues to be attractive given its high level of explainability.

\section{Lessons from Retrospection}\label{Sec:different}

Having focused on some of the more positive aspects of the research in our 1997 paper, I now turn to those things that, with hindsight, should or could have been done differently.  These, in the main, relate to (i) analogy-based project effort prediction and (ii) the actual conduct of the experiment and its reporting.

First then, topics that could have been fruitfully explored.
\begin{itemize}
    \item Explore alternative distance metrics and handle categorical variables better, especially when the variable has many levels, for instance, country.  One possibility would be to investigate hybrid models, e.g., using decision trees or rule-based systems to incorporate categorical variables directly into decision trees or rule-based methods instead of reducing them to binary values. This approach naturally handles non-linear and hierarchical relationships.
    \item Use of different case representations to the our standard approach of feature vectors, e.g., hierarchies and networks to capture the richer structure of the data
    \item Enhanced modelling of uncertainty both in terms of the individual project features i.e., measurement error and in terms of donor dissimilarity from the target project.  A side effect of standardised Euclidean distance $(x_i-min(x))/(max(x)-min(x))$ is that an outlier will rescale all distances and compress the remaining cases in the feature space.
\end{itemize}

Second, and of wider relevance, is the design and conduct of our empirical study.
\begin{itemize}
    \item Upon reflection, it is undeniable we had a preferred outcome and this bias potentially manifested itself in subtle ways that influenced reporting and where we had flexibility, it might have changed our analysis \cite{Stapor2021design}.  Solutions include differentiating between exploratory and confirmatory analysis, pre-registration \cite{Bakker2020ensuring} and blind analysis \cite{Fucci2016external}.  However, to be absolutely clear, such biases were subconscious, and in no way did either author intentionally distort the analysis or misrepresent the results. 
    \item We also focused on the best result without reporting variance or distribution of the predictive performances possible \cite{Jedlitschka2005reporting}. Any potential user of a prediction system, particularly when high value decisions are being made as is the case with software projects, will want to know the variability of prediction accuracy to complement typical, or less helpfully, best case accuracy.
    \item The next area of the study where I consider that there is room for improvement relates to the inferencing mechanism.  Although it was not explicit, in the 1997 paper, we simply used a better-than argument for the two accuracy metrics for analogy and stepwise regression. Since there were nine data sets and therefore $9 \times 2$ comparisons we deployed something akin to vote counting.  
    
    Subsequently, the idea has been popularised of null hypothesis significance testing (NHST) based on comparing p-values with some pre-determined threshold, almost universally $\alpha=0.05$ \cite{Jorgensen2001impact,Wallshein2015software}.  Despite this widespread adoption of NHST inferencing, there are some substantial drawbacks.  It has long been pointed out by the statistics community, that p-values do not convey the information that is usually desired, namely the likelihood of the alternate hypothesis given the observed data \cite{Berkson1942tests,Gardner1986confidence,Mcshane2019abandon}. 
    Put simply users of NHST would like $P(H|D)$ but unfortunately what we have is $P(D|H)$ that is the long run probability P of seeing data D as or more extreme, \textit{given} that the null hypothesis H is true.  But the difficulty is the null hypothesis is almost invariably some kind of straw man that nobody believes leading to the incongruous situation where P can only be interpreted in this way given the one thing we do not believe is true!  Another problem it forces the dichotomous interpretation that a result is either true or not true.  However, with for example project effort prediction systems, we would like to know how much better P1 is over P2.  A significant result says nothing about the size of the effect.
    \item Our choice of accuracy metrics, although noting some problems with MMRE and Pred(25) we in the end chose to use them because they were widely used by other researchers.  This was not a strong argument and can result in a research area becoming mired in the past.  Subsequently, we adopted a more principled position \cite{Kitchenham2001accuracy} and proposed sound alternatives such Standardised Accuracy \cite{Shepperd2012evaluating}.
    \item We reported all our results in terms of tables of prediction accuracy metrics without providing a clear idea of the real world effect.  It would have been more useful to report effect sizes along with confidence intervals \cite{Ellis2010essential} which can easily computed using a bootstrap procedure even when assumptions regarding normality, equality of variances, etc are hard to justify \cite{Manly2018randomization}.  Frequently standardised effect sizes are used to enable easier interpretation and comparison between different data sets and metrics.  With hindsight reporting results as standardised effects such as the Cohen's $d$, i.e., $(\overline{m_1} - \overline{m_2} )/s$ where $m_1$ and $m_2$ are the two treatments being compared and $s$ is some estimate of the pooled standard deviations \cite{Ellis2010essential}.  
    \item Another concern is there was no explicit reporting of threats to validity \cite{Shadish2002,Wohlin2012experimentation}.  This is unhelpful to readers attempting to judge the significance of the results and also to researchers seeking to follow up and improve upon the empirical research.  Shadish, Cook and Campbell emphasise the importance of addressing threats comprehensively to ensure credible causal inferences. They categorise validity into four types: internal, external, construct and statistical conclusion validity, each susceptible to specific threats.  Transparent reporting allows readers to critically evaluate the robustness and limitations of the study’s conclusions, promoting scientific rigour and reproducibility, even if there is a risk this can degenerate into a box ticking exercise \cite{Lago2024threats}.  
\item Finally, and rather importantly, our study is not now reproducible\footnote{Note, we define reproducibility as repeating a study as faithfully as possible with the goal of determining whether there have been errors in commission and/or reporting. This is the \textit{sine qua non} of science since if we can't reliably repeat an experiment then research progress becomes impossible.  In contrast, the goal of replication is addressing confidence in and the generalisability of the results.} despite our attempts, at the time, to share the data and code.  Unfortunately an ad hoc approach to hosting data and transient url's has led to the subsequent, non-availability of some data sets.  More problematic, the choice of Visual Basic on Windows NT/95 with bought-in VBX controls means that the CBR tool Angel cannot now be executed.  A newer version was produced, written in Java but that introduced some changes and refinements such that we cannot exactly reproduce the results of the 1997 paper.   

Encouraginly, the Open Science movement \cite{Munafo2017manifesto} is gathering momentum and apart from better awareness, we are now being provided with better tools, enduring platforms such zenodo\footnote{\url{https://zenodo.org }}, figshare\footnote{\url{https://figshare.com}} and the Open Science Framework.\footnote{\url{https://osf.io }}  Other useful infrastructure includes provision of Digital Object Identifiers (DOIs) and Docker containers, that enable researchers to package and share reproducible, version-controlled environments for experiments and analyses, ensuring consistency across different systems and hopefully more effective future-proofing.
\end{itemize} 
%    Shull, F., Carver, J., Vegas, S., & Juristo, N. (2008). The role of replications in empirical software engineering. Empirical Software Engineering, 13(2), 211--218. 
%    Ince, D., Hatton, L., & Graham-Cumming, J. (2012). The case for open computer programs. Nature, 482, 485–488. https://doi.org/0.1038/nature10836 
%Ioannidis, J. (2014). How to Make More Published Research True. PLoS Medicine, 11(10), e1001747. 

\section{Lessons for the future}\label{Sec:lessons}

Software engineering has changed greatly in the past 28 years.  The idea of non-trivial, coherent software projects is less ubiquitous and there has been a shift to Agile methodologies.  Effort estimation challenges have evolved with Agile's iterative development cycles and less predictable scope.  In parallel, there is growing diversity of software domains, for example the emergence of AI, mobile apps and cloud services.  These again introduce new challenges in effort prediction not addressed in traditional datasets.

More enduring than analogy-based project effort prediction are some of the methodological approaches embodied in our 1997 study.  These include the idea of robust empirical validation using multiple data sets against meaningful benchmarks and the sharing and publishing of data sets and tools.  Our focus on adequate out-of-sample test sets and loss functions remains fundamental.  

Of course, there are many aspects that were we to repeat the study today, we would do differently.  These are described in some detail in Section~\ref{Sec:different}, however, the key points relate to the Open Science movement and the need to give adequate support for study reproducibility especially over time.  Better and completer reporting is the other key area that needs to more adequately adopted.  It is important to address the variability of results as well as best or typical results. Likewise, threats to validity also need to be tackled meaningfully and reported in depth.

Nonetheless, when reflecting back over the past 28 years, it is a privilege to have played some small role in the development of prediction research within the domain of software engineering.  I look forward to the next 28 years!

\section*{Acknowledgment}

Martin Shepperd would like to thank Chris Schofield for the development of the initial Visual Basic version of the ANGEL estimation tool that underpins much of the research reported in the original paper.  

\bibliographystyle{abbrv}
\balance
\bibliography{TSE_Retro_2025}

\begin{thebibliography}{10}

\bibitem{Bakker2020ensuring}
M.~Bakker, C.~Veldkamp, M.~van Assen, E.~Crompvoets, H.~Ong, B.~Nosek, C.~Soderberg, D.~Mellor, and J.~Wicherts.
\newblock Ensuring the quality and specificity of preregistrations.
\newblock {\em PLoS Biology}, 18(12):e3000937, 2020.

\bibitem{Bergmann2005representation}
R.~Bergmann, J.~Kolodner, and E.~Plaza.
\newblock Representation in case-based reasoning.
\newblock {\em The Knowledge Engineering Review}, 20(3):209--213, 2005.

\bibitem{Berkson1942tests}
J.~Berkson.
\newblock Tests of significance considered as evidence.
\newblock {\em Journal of the American Statistical Association}, 37(219):325--335, 1942.

\bibitem{Boehm1984software}
B.~Boehm.
\newblock Software engineering economics.
\newblock {\em IEEE Transactions on Software Engineering}, 10(1):4--21, 1984.

\bibitem{Conte1985software}
S.~Conte, H.~Dunsmore, and V.~Shen.
\newblock Software effort estimation and productivity.
\newblock In M.~Yovits, editor, {\em Advances in Computers}, volume~24, pages 1--60. Elsevier, 1985.

\bibitem{conte1986software}
S.~Conte, H.~Dunsmore, and V.~Shen.
\newblock {\em Software engineering metrics and models}.
\newblock Benjamin-Cummings Publishing Co., Inc., 1986.

\bibitem{Ellis2010essential}
P.~Ellis.
\newblock {\em The essential guide to effect sizes: Statistical power, meta-analysis, and the interpretation of research results}.
\newblock Cambridge University Press, 2010.

\bibitem{Fucci2016external}
D.~Fucci, G.~Scanniello, S.~Romano, M.~Shepperd, B.~Sigweni, F.~Uyaguari, B.~Turhan, N.~Juristo, and M.~Oivo.
\newblock An external replication on the effects of test-driven development using a multi-site blind analysis approach.
\newblock In {\em 10th ACM/IEEE International Symposium on Empirical Software Engineering and Measurement}, pages 1--10, 2016.

\bibitem{Gardner1986confidence}
M.~Gardner and D.~Altman.
\newblock Confidence intervals rather than p values: estimation rather than hypothesis testing.
\newblock {\em British Medical Journal}, 292(6522):746--750, 1986.

\bibitem{Idri2015analogy}
A.~Idri, F.~Azzahra~A., and A.~Abran.
\newblock Analogy-based software development effort estimation: A systematic mapping and review.
\newblock {\em Information and Software Technology}, 58:206--230, 2015.

\bibitem{Idri2016improved}
A.~Idri, M.~Hosni, and A.~Abran.
\newblock Improved estimation of software development effort using classical and fuzzy analogy ensembles.
\newblock {\em Applied Soft Computing}, 49:990--1019, 2016.

\bibitem{Jedlitschka2005reporting}
A.~Jedlitschka and D.~Pfahl.
\newblock Reporting guidelines for controlled experiments in software engineering.
\newblock In {\em International Symposium on Empirical Software Engineering, (ISESE 2005)}. IEEE, 2005.

\bibitem{John1994irrelevant}
G.~John, R.~Kohavi, and K.~Pfleger.
\newblock Irrelevant features and the subset selection problem.
\newblock In {\em 11th International Machine Learning Conference, 1994}, pages 121--129. Morgan-Kaufmann, 1994.

\bibitem{Jorgensen2001impact}
M.~J{\o}rgensen and D.~Sj{\o}berg.
\newblock Impact of effort estimates on software project work.
\newblock {\em Information and software technology}, 43(15):939--948, 2001.

\bibitem{Kaushik2022stacking}
A.~Kaushik, P.~Kaur, N.~Choudhary, and Priyanka.
\newblock Stacking regularization in analogy-based software effort estimation.
\newblock {\em Soft Computing}, pages 1--20, 2022.

\bibitem{Kemerer1987empirical}
C.~Kemerer.
\newblock An empirical validation of software cost estimation models.
\newblock {\em Communications of the ACM}, 30(5):416--429, 1987.

\bibitem{Keung2008analogy}
J.~Keung, B.~Kitchenham, and R.~Jeffery.
\newblock Analogy-x: providing statistical inference to analogy-based software cost estimation.
\newblock {\em IEEE Transactions on Software Engineering}, 34(4):471--484, 2008.

\bibitem{Kirsopp2003empirical}
C.~Kirsopp, E.~Mendes, R.~Premraj, and M.~Shepperd.
\newblock An empirical analysis of linear adaptation techniques for case-based prediction.
\newblock In {\em 5th International Conference on Case-Based Reasoning (ICCBR 2003)}, pages 231--245. Springer, 2003.

\bibitem{kirsopp2002search}
C.~Kirsopp, M.~Shepperd, and J.~Hart.
\newblock Search heuristics, case-based reasoning and software project effort prediction.
\newblock pages 1367--1374, 2002.

\bibitem{Kitchenham2001accuracy}
B.~Kitchenham, L.~Pickard, S.~MacDonell, and M.~Shepperd.
\newblock What accuracy statistics really measure.
\newblock {\em IEE Proceedings-Software}, 148(3):81--85, 2001.

\bibitem{Kitchenham1985software}
B.~Kitchenham and N.~Taylor.
\newblock Software project development cost estimation.
\newblock {\em Journal of Systems and Software}, 5(4):267--278, 1985.

\bibitem{Kocaguneli2011exploiting}
E.~Kocaguneli, T.~Menzies, A.~Bener, and J.~Keung.
\newblock Exploiting the essential assumptions of analogy-based effort estimation.
\newblock {\em IEEE Transactions on Software Engineering}, 38(2):425--438, 2011.

\bibitem{Kolodner1992introduction}
J.~Kolodner.
\newblock An introduction to case-based reasoning.
\newblock {\em Artificial Intelligence Review}, 6(1):3--34, 1992.

\bibitem{Kraut1995coordination}
R.~Kraut and L.~Streeter.
\newblock Coordination in software development.
\newblock {\em Communications of the ACM}, 38(3):69--82, 1995.

\bibitem{Lago2024threats}
P.~Lago, P.~Runeson, Q.~Song, and R.~Verdecchia.
\newblock Threats to validity in software engineering--hypocritical paper section or essential analysis?
\newblock In {\em The 18th ACM/IEEE International Symposium on Empirical Software Engineering and Measurement}, pages 314--324, 2024.

\bibitem{Lederer1995causes}
A.~Lederer and J.~Prasad.
\newblock Causes of inaccurate software development cost estimates.
\newblock {\em Journal of Systems and Software}, 31(2):125--134, 1995.

\bibitem{Li2009study}
Y.~Li, M.~Xie, and T.~Goh.
\newblock A study of project selection and feature weighting for analogy based software cost estimation.
\newblock {\em Journal of systems and software}, 82(2):241--252, 2009.

\bibitem{Mair2005consistency}
C.~Mair and M.~Shepperd.
\newblock The consistency of empirical comparisons of regression and analogy-based software project cost prediction.
\newblock In {\em 2005 International Symposium on Empirical Software Engineering (ISESE)}, 2005.

\bibitem{Manly2018randomization}
B.~Manly.
\newblock {\em Randomization, bootstrap and Monte Carlo methods in biology}.
\newblock chapman and Hall/CRC, 2018.

\bibitem{Mcshane2019abandon}
B.~McShane, D.~Gal, A.~Gelman, C.~Robert, and J.~Tackett.
\newblock Abandon statistical significance.
\newblock {\em The American Statistician}, 73:235--245, 2019.

\bibitem{Munafo2017manifesto}
M.~Munaf{\`o}, B.~Nosek, D.~Bishop, K.~Button, C.~Chambers, N.~Percie~du Sert, U.~Simonsohn, E.~Wagenmakers, J.~Ware, and J.~Ioannidis.
\newblock A manifesto for reproducible science.
\newblock {\em Nature human behaviour}, 1(1):1--9, 2017.

\bibitem{Phannachitta2017stability}
P.~Phannachitta, J.~Keung, A.~Monden, and K.~Matsumoto.
\newblock A stability assessment of solution adaptation techniques for analogy-based software effort estimation.
\newblock {\em Empirical Software Engineering}, 22:474--504, 2017.

\bibitem{Prietula1996software}
M.~Prietula, S.~Vicinanza, and T.~Mukhopadhyay.
\newblock Software-effort estimation with a case-based reasoner.
\newblock {\em Journal of Experimental \& Theoretical Artificial Intelligence}, 8(3-4):341--363, 1996.

\bibitem{Shadish2002}
W.~Shadish, T.~Cook, and D.~Campbell.
\newblock {\em Experimental and quasi-experimental designs for generalized causal inference}.
\newblock Houghton Mifflin, Boston, 2002.

\bibitem{Shepperd2012evaluating}
M.~Shepperd and S.~MacDonell.
\newblock Evaluating prediction systems in software project estimation.
\newblock {\em Information and Software Technology}, 54(8):820--827, 2012.

\bibitem{Shepperd1997estimating}
M.~Shepperd and C.~Schofield.
\newblock Estimating software project effort using analogies.
\newblock {\em IEEE Transactions on Software Engineering}, 23(11):736--743, 1997.

\bibitem{Song2011predicting}
Q.~Song and M.~Shepperd.
\newblock Predicting software project effort: A grey relational analysis based method.
\newblock {\em Expert Systems with Applications}, 38(6):7302--7316, 2011.

\bibitem{Song2005using}
Q.~Song, M.~Shepperd, and C.~Mair.
\newblock Using grey relational analysis to predict software effort with small data sets.
\newblock In {\em 11th IEEE International Software Metrics Symposium (METRICS'05)}, pages 10--pp. IEEE, 2005.

\bibitem{Stapor2021design}
K.~Stapor, P.~Ksieniewicz, S.~Garc{\'\i}a, and M.~Wo{\'z}niak.
\newblock How to design the fair experimental classifier evaluation.
\newblock {\em Applied Soft Computing}, 104:107219, 2021.

\bibitem{Wallshein2015software}
C.~Wallshein and A.~Loerch.
\newblock Software cost estimating for cmmi level 5 developers.
\newblock {\em Journal of Systems and Software}, 105:72--78, 2015.

\bibitem{Wohlin2012experimentation}
C.~Wohlin, P.~Runeson, M.~H{\"o}st, M.~Ohlsson, B.~Regnell, and A.~Wessl{\'e}n.
\newblock {\em Experimentation in software engineering}, volume 236.
\newblock Springer, 2012.

\end{thebibliography}

\end{document}